\documentstyle[11pt,moriond,epsfig]{article}

\def\be{\begin{equation}}
\def\ee{\end{equation}}
\def\bea{\begin{eqnarray}}
\def\eea{\end{eqnarray}}

\def\alt{\stackrel{<}{_\sim}}
\def\agt{\stackrel{>}{_\sim}}
\def\apropto{\stackrel{\propto}{_\sim}}
\def\ord#1{{\cal O}(#1)}
\begin{document}
\title{COULOMB BLOCKADE FLUCTUATIONS IN DISORDERED QUANTUM DOTS}

\author{P. N. WALKER$^1$, G. MONTAMBAUX$^2$ and Y. GEFEN$^3$}

\address{$^1$Department of Physics and Astronomy, University College London,
Gower St., London, England.\\ $^2$Laboratoire de Physique des Solides,
associ\'e au CNRS, Universit\'{e} Paris--Sud, 91405 Orsay, France.\\
$^3$Department of Condensed Matter Physics, The Weizmann Institute
of Science, 76100 Rehovot, Israel.}

\maketitle\abstracts{We discuss some recent results on the statistics of
the Coulomb Blockade in disordered quantum dots containing spinless
interacting fermions using the self-consistent Hartree-Fock
approximation. We concentrate on the regime $r_s\agt 1$, with finite 
dimensionless conductance $g$. We present significantly different results
for the cases of a Coulomb and a nearest-neighbour bare interaction.
We show that neglecting rearrangements when the particle number is changed
(Koopmans' approximation) can lead to large errors, and show how the spectral
statistics of the self-consistent single particle orbitals evolve with the
interaction strength.}

\section{Introduction}

The Coulomb Blockade to tunnelling in quantum dots, where the charge in the
dot is a good quantum number, provides an experimentally accessible measure
of the difference between ground state energies of successive total particle
number \cite{Sivan1,Simmel1,Marcus1,Simmel2,Ashoori1,Ashoori2}. More
specifically, the accessible quantity is the second difference of ground state
energies $E_G(N,V_G)$:
\begin{equation}
\Delta_2(N)= E_G(N+1,V_G)-E_G(N,V_G)-E_G(N,V'_G) + E_G(N-1,V'_G)\ ,
\label{D2VG}
\end{equation}
where $V_G$ parametrises the gate voltage environment at each resonant
tunnelling point.
Within the constant interaction (CI) model the ground state energy
is simply the sum of filled single particle energies ${\cal E}(n)$ plus
$N(N-1)e^2/2C$, where $e^2/C$ is the constant interaction, and
$C\propto L^{-1}$ is the geometric capacitance. Taking
$V_G=V'_G$, the peak spacing trivially reduces to
\begin{equation}
\Delta_2(N)= {\cal E}(N+1)-{\cal E}(N)+e^2/C
\label{V0}
\end{equation}
and so, in the diffusive regime, displays Wigner-Dyson (WD)
statistics \cite{Mehta} shifted by $e^2/C$,
up to corrections in one over the dimensionless conductance, $g$.

Recent experiments on quantum dots \cite{Sivan1,Simmel1,Marcus1,Simmel2}
have shown that whilst the mean peak spacings are apparently well described by
the CI model, the fluctuations are not described by Wigner-Dyson statistics.
It is found that the distribution of $\Delta_2$ is roughly Gaussian 
\cite{Sivan1,Marcus1,Simmel2}, with broader non-Gaussian tails seen in
Ref. 3. In Refs. 1,2,4 the variance of the
fluctuations was found to be considerably larger than that given by the WD
distribution. Exact numerical calculations on very small systems
\cite{Sivan1} also exhibit Gaussian fluctuations on a scale much larger
than $\Delta\propto L^{-2}$, and it was suggested that the typical size
of the fluctuations scale universally with the average spacing
$\langle\Delta_2\rangle \apropto e^2/C$.

In an attempt to bridge the gap between theory \cite{BerkAlt1} and experiment, 
Blanter {\it et al} \cite{Blanter97}
evaluated the fluctuations within a Hartree-Fock framework, neglecting
effects due to the change in gate voltage $V_G$ to $V'_G$.
To this end they applied the Random Phase Approximation (RPA) to generate
the screened interaction in the confined geometry, and assumed
that all HF level spacings, except for the Coulomb gap, are described
by WD statistics. Implicitly assuming Koopmans' theorem \cite{Koopmans} to
be valid (which neglects single-particle rearrangements when the particle
number distribution is altered), and using wavefunction statistics
established for non-interacting electrons in a random potential,
they calculated the fluctuations of $\Delta_2$ beyond the CI model.
These additional fluctuations were found to be
parametrically small (in $1/g$) and proportional to $\Delta$.
Hence, the total fluctuations in $\Delta_2$ were found to be proportional
to $\Delta$. The disagreement between experiment and theory therefore remained.
The analysis of Ref. 9 is consistent in
the limits $g\gg 1$, $r_s\ll 1$. The parameter $r_s$ characterises the
relative importance of interactions in the electronic system, and
is defined as the mean electron separation in units of the effective
Bohr radius. Typical values of $r_s\sim\ord{1}$ have been reported in
experiment \cite{Sivan1,Simmel2}.

To understand why theory and exact numerics are in disagreement, we consider
an effective single particle model as in Ref. 9, but
generate the single particle orbitals self-consistently using the Hartree-Fock
approximation (SCHF). This model allows for the rearrangements when particles
are added or removed which was neglected in Ref. 9, and does
not require any conjecture about the wavefunction correlations (the two
issues are of course intimately linked). We are also able to consider
much  larger samples than is feasible by exact
diagonalisation \cite{Sivan1}. This approximation has also been seen to
be quite good for the calculation of persistent currents in similar
systems \cite{Poilblanc}.
We show that fluctuations large compared to the single particle level spacing
can arise without recourse to varying the sample shape, size or gate-to-dot
coupling, supposing these to be {\it additional} effects. We further
demonstrate that approximating the addition spectrum spacings by applying
Koopmans' theorem can lead to large errors in the calculation of the
spacing statistics. A more detailed presentation of these results can be
found elsewhere \cite{Walker2,Walker1}.

%
\section{The Model}
\label{Model}
%
We consider the following tight binding Hamiltonian for spinless
fermions
\begin{equation}H = \sum_i w_i c^+_i c_i - 
t\sum_{i, \eta}c^+_{i+\eta} c_{i} +
{U\over 2} \sum_{ij} M_{ijij} c^+_i c^+_j c_j c_i
\label{H}
\end{equation}
where $i$ is the site index,
$\eta$ describes the set of nearest neighbours, $w_i$ is the
random on site energy in the range $[-W/2,W/2]$, and
$t$ the hopping matrix element, henceforth taken as unity. We study
separately, both a Coulomb interaction potential,
\begin{equation}
M_{ijij}= 1/|{\bf r}_i-{\bf r}_j|\ ,
\label{lrmel}
\end{equation}
and a short range potential,
\begin{equation}
M_{ijij}=\delta_{i,i+\eta}\ ,
\label{srmel}
\end{equation}
there is no need to add a constant interaction as it does not affect the
physics.

We consider a 2D system with periodic boundary conditions, and choose to
define 
\begin{equation}
|{\bf r}_i-{\bf r}_j|^2 \equiv [L_x^2\sin^2 (\pi n_x /L_x)+
L_y^2\sin^2(\pi n_y/L_y)]/\pi^2\ ,
\end{equation}
where $(n_x,n_y) \equiv {\bf r}_i-{\bf r}_j$, and $L_x$, $L_y$ are the
sample dimensions.

All distances are measured in units of the lattice spacing $a$; the
physical parameters are therefore $U=e^2/a$, $t=\hbar^2/2ma^2$.
The standard definition for $r_s$ is given,
for low filling, by $r_s=U/(t\sqrt{4\pi\nu})$, where $\nu=N/A$ is the
filling factor on the tight binding lattice with $A$ sites.
The dimensionless conductance $g$ can be approximated, again for low
filling, using the Born approximation. We find
$g= 96\pi\nu(t/W)^2$, which is valid for $A,N\gg g\gg 1$.

%
\section{The SCHF Approximation}
\label{SCHF}
%
There are a few aspects of this approximation scheme which warrant discussion
before presenting the results. Firstly we should stress that this scheme is
non-perturbative, but uncontrolled. To ensure that the converged solutions
obtained were at least local minima, and hence the RPA excitations
stable \cite{Thouless1}, we diagonalised the total energy
curvature matrix \cite{Thouless2} in some representative samples. The
calculational effort involved is too heavy to do this for all cases.

The issue of which bare potential (\ref{lrmel},\ref{srmel}) should be used is
subtle. If screening is generated externally (by close metallic plates for
example), it is reasonable to insert a short-ranged interaction for all
interaction strengths. It seems \cite{Walker1} that this model may then
be useful to model the quantum dots of Refs. 5,6.
It is not clear however that this form should be
taken to represent screening by high-energy particle hole pairs (the RPA
approximation). The reason being that the self-consistency process
automatically incorporates much of the physics included in the RPA process.
In other words, it generates many of the diagrams included in the RPA
calculation, and so leads to double counting. This problem does not
arise if one applies the Coulomb bare potential. However, we know from
basic considerations that the exchange (Fock) interaction should be
screened, and the SCHF process is unable to achieve this. In clean systems
this can lead to large errors, here however, we have verified that the typical
fluctuations of the exchange contribution are smaller than those of the
Hartree contribution over the parameter range considered. This
suggests that the error made in not screening the exchange term correctly is
not overly important.
We expect then, that the results for a long ranged interaction should be
representative of the exact numerical calculations.

The simplest way to interpret the RPA calculation, is the (average) response
of the electron gas to the insertion of a test charge $e$. Since the excess
charge cannot be pushed to infinity in a bounded dot, it is perhaps not
surprising that an explicit calculation of the RPA screened interaction
shows that a charge $e$ accumulates on the boundary \cite{Blanter97}
\footnote{We note that if the dot is supposed to be charge neutral, it may
be more consistent if the electron gas were to see an excess background
charge of $-e$, plus the {\sl test} electron in question}.
It was found in Ref. 9
that such {\sl boundary} effects could dominate the interaction induced
fluctuations of $\Delta_2$. In any event, the effect of the boundaries
is geometry specific; we remove it by considering a torus geometry.

Since the SCHF approximation attempts to generate static screening by
rearranging the effective-single-particle orbitals, there must be a
different interpretation between the SCHF approximation with a Coulomb
bare potential, and HF with RPA. In the latter,
most charge rearrangements are made via the RPA summation, and as such, are
separated from the effective-single particles. In our time-independent
self-consistent theory, it is not possible to make such a separation,
and indeed the job of charge rearrangement is carried purely by the
effective-single-particle orbitals. Clearly then, the meaning of the
effective-single-particle is not the same in the two cases.

Since the validity of Koopmans' theorem is analysed here, we explain our
motivation for investigating an effect that is second order effect in $r_s$.
At sufficiently small $r_s$, where non-interacting wavefunction correlations
can be employed, the effects of rearrangements are negligible. It is often
assumed that they are small for the further reason that adding a particle
generates only $\ord {N}$ corrections to the existing
single-particle wavefunctions. We stress however, that when evaluating
the ground-state energy, one must sum over $N$ such corrections (as
well as $N^2$ of $\ord {N^2}$), and so the cumulative effect can
be large. When $r_s$ is sufficiently large as to affect the wavefunction
correlations, it is further plausible, that the $\ord {r_s^2}$ corrections
to the wavefunctions correlate with the unperturbed wavefunctions to
generate typical fluctuations of order $r_s$.

\section{Nearest-Neighbour Interactions}

In Ref. 11 we show that the distribution of level spacings
for both occupied and unoccupied states evolve with $r_s$ from
a Wigner-Dyson distribution towards a Poisson-like distribution. This
is more pronounced for the unoccupied states, which only have a meaning if
Koopmans' approximation is invoked. This tendency is rather weak however,
and for $r_s\sim\ord {1}$, the distribution is almost indistinguishable from
Wigner-Dyson. On the other hand, we found that the self-consistently
calculated gap ($\Delta_2$) evolved comparatively rapidly towards a Gaussian
distribution, and was essentially Gaussian for $r_s \agt 2$.
The main difficulty with observing the typical interaction
induced fluctuations for $r_s \alt 1$ was that they were small compared
to $\Delta$. In this regime we confirmed that they scaled like
$r_s \Delta$ for fixed $g$, in agreement with Ref. 9. Indeed,
if the wavefunctions remain ergodic, it is clear that this scaling must hold.
It was too difficult to identify the disorder scaling with any certainty.
In this regime, applying Koopmans' theorem did of course generate errors,
but they were found to be relatively small: the mean error scaled like
$r_s^2 \Delta$, and the error in the typical fluctuations was too small
to analyse with confidence.

When $r_s\agt 1$, we found a dramatic development in the SCHF ground state,
and a correspondingly dramatic growth in the error incurred through Koopmans'
approximation. We also noted that the system size scaling of the mean
Koopmans' error only settled to the $r_s^2 \Delta$ dependence for the larger
systems considered ($A \agt 50$ lattice sites); well beyond the current
limitations of exact calculations.
Whilst the single-particle wavefunctions remained essentially
ergodic, incipient charge density modulations (CDMs) led to an unusual, and
highly non-universal fluctuation behaviour. In this regime, deviations in
$\langle\Delta_2\rangle$ from the CI model prediction were observed, and
identified to be $\ord {U/A}$. They were shown to be a direct consequence of
the incipient CDMs. These CDMs are analysed in much greater detail in
Ref. 13.

In conclusion, we find that for weak interactions, the typical fluctuations
$\delta\Delta_2 \sim r_s\Delta$ for fixed $g>1$, in broad agreement with 
Blanter {\it et al} \cite{Blanter97}, but not with exact
numerics \cite{Sivan1}. For $r_s\agt 1$,
incipient charge density modulations cause a significant change in the
statistics of $\Delta_2$. It is suggested in Ref. 13, that such
effects may be relevant to the unusual addition spectra observed in the
capacitance experiments of Refs. 5,6.

\section{Coulomb Interactions}

As for the nearest-neighbour interaction case, the distribution of level
spacings for both occupied and unoccupied states were investigated in
Ref. 11. They were both found to remain very close to the WD
distribution for all interaction strengths considered (up to $r_s \sim 5$).
The gap ($\Delta_2$) distribution was found to evolve towards a Gaussian
distribution much more rapidly than changes in the above level spacings,
but the scale on which this happened depended on the size of the sample.

The mean error generated by Koopmans' approximation was found to tend towards
a simple quadratic dependence on $r_s$ as the system size was increased to the
maximum size considered of 144 sites and 37 particles. As for the nearest
neighbour case, the system size scaling changed around $A\approx 50$ lattice
sites. The mean Koopmans' error therefore
diverges with system size when compared to both $\langle \Delta_2\rangle$
and $\Delta$. 
The Koopmans' approximation error in the typical fluctuations was found to
become very large ($\ord {\delta\Delta_2} $), no attempt was made to generate
an empirical formula to describe it.

The divergent relative error of Koopmans' approximation to the mean charging
energy was one of the more striking results of this work. Whilst the CI
model worked remarkably well, a supposed improvement on it (Koopmans'
theorem about a self-consistent basis) was found to be worse, and moreover,
could only be made meaningful for $r_s < \ord {1/L}$. This contradicts
the expectation that Koopmans' theorem should become a good approximation
for $r_s \ll 1$, regardless of the system size.
In the CI model this error vanishes identically,
while in Blanter {\it et al} \cite{Blanter97}, the terms which
generate it were implicitly assumed to vanish.

Turning to the main object of investigation, the SCHF approximation was
found to yield typical fluctuations in the charging energy of
$\delta\Delta_2 \sim 0.52\Delta+
a\langle\Delta_2\rangle/\sqrt{g} +\ord {r_s^2}$,
where $a$ is a constant. A similar result was also found independently by
Bonci and Berkovits \cite{Bonci}. Whilst this result disagrees with the claim
in Ref. 1 that the fluctuations are universal, the system
size scaling is consistent. The $\ord{r_s^2}$ contribution arises due to
interaction induced correlations between the single-particle wavefunctions,
and eventually dominates the fluctuations as $r_s$ is increased. However,
the validity of the SCHF approximation becomes more questionable as $r_s$
is increased, and it is not clear to what extent the results are valid
in this strong interaction regime. One might
expect that the combined fluctuations of the single particle orbitals and
the RPA-bosons in the HF with RPA model are similar to the fluctuations
generated in the Coulombic SCHF model. Since we do not find agreement with
Blanter {\it et al} \cite{Blanter97}, this appears not to be the case.
A direct calculation of fluctuations in the polarisability
\cite{BerkAlt2,Blanter97} does not account for this discrepancy.
It would appear that either screening is not generated in our system
(or indeed in the exact calculations on very small samples), or that it
is not correct to separate out the RPA-bosons
from the fluctuation effects in the the usual way.

In summary, we found that Koopmans theorem fails dramatically in the case
of a Coulomb bare interaction. We find that the typical interaction induced
fluctuations are proportional to $\langle\Delta_2\rangle$ to leading order
in $r_s$, but they are not universal, rather they vanish with the
dimensionless conductance like $1/\sqrt{g}$. Other related work
has been independently carried out by Levit and
Orgad \cite{Levit}, Cohen, Richter and Berkovits \cite{Cohen}, and
Bonci and Berkovits \cite{Bonci}.

\section*{Acknowledgments}

We acknowledge discussions with Ya. Blanter, A. Finkelstein, S. Levit,
C. Marcus, A. Mirlin, D. Orgad, H. Orland, F. von Oppen, F. Piechon,
B. Shklovskii and U. Sivan.
We acknowledge support from the EU TMR fellowship ERBFMICT961202,
the German-Israeli Foundation, the U.S.-Israel Binational-Science
Foundation and the Minerva Foundation.
Much of the numerical work was performed using IDRIS facilities.

\section*{References}

\end{document}